\documentclass[a4paper,aps,prl,twocolumn,groupedaddress,showkeys,showpacs,
  floats,floatats,floatfix]{revtex4}
  \usepackage{CJK}
\usepackage[utf8]{inputenc}
\usepackage[english]{babel}
\usepackage{graphicx}
\usepackage{dcolumn} 
\usepackage{bm}
\usepackage{natbib}
\usepackage{latexsym}
\usepackage{mathrsfs}
\usepackage{amssymb}
\usepackage{amsmath}
\usepackage{amscd}
\usepackage{color}
\usepackage{pifont}
\usepackage{pstricks,pst-node,pst-text,pst-3d}
\usepackage{verbatim}
\usepackage{ulem}
\usepackage[T1]{fontenc}
\usepackage{hyperref}
\hypersetup{
  colorlinks   = true, 
  urlcolor     = black, 
  linkcolor    = red, 
  citecolor    = blue 
}
\definecolor{aquamarine}{rgb}{0.5, 1.0, 0.83}
\definecolor{blue-violet}{rgb}{0.54, 0.17, 0.89}

\newcommand{\cbeims}[1]{{\color{red}#1}}

\newrgbcolor{Red}{1.0 0.0 1.0}

\begin{document}
\title{Finite-time Lyapunov  fluctuations and the upper bound of classical and quantum out-of-time-ordered expansion rate exponents}

\author{Miguel A. Prado Reynoso}
\email{prado_angel92@hotmail.com}
\affiliation{Instituto de Ciencias Físicas, Universidad Nacional Autónoma de México, 62210, Cuernavaca, Morelos, Mexico}

\author{Guilherme J. Delben}
\email{guilherme.delben@ufsc.br}
\affiliation{Departamento de Ciências Naturais e Sociais, Universidade Federal de Santa Catarina, 89520-000 Curitibanos, Brazil}

\author{Martin Schlesinger}
\affiliation{TraceTronic GmbH, Stuttgarter Str. 3, 01189 Dresden, Germany}
\email{martin-schlesinger@gmx.de}


\author{Marcus W.~Beims}
\email{mbeims@fisica.ufpr.br}
\affiliation{Departamento de F\'\i sica, Universidade Federal do Paran\'a, 
81531-980 Curitiba, Paraná, Brazil}

\date{\today}
\begin{abstract}
This Letter demonstrates for chaotic maps (logistic, classical and quantum standard maps (SMs)) that the exponential growth rate ($\Lambda$) of the out-of-time-ordered four-point correlator (OTOC) is equal to the classical Lyapunov exponent ($\lambda$)  \textit{plus} fluctuations ($\Delta^{\mbox{\tiny (fluc)}}$) of the one-step finite-time Lyapunov exponents (FTLEs). Jensen's inequality provides the upper bound $\lambda\le\Lambda$  for the considered systems. Equality  is restored with $\Lambda  = \lambda + \Delta^{\mbox{\tiny (fluc)}}$, where $\Delta^{\mbox{\tiny (fluc)}}$ is quantified by $k$-higher-order cumulants of the FTLEs. Exact expressions for $\Lambda$ are derived and numerical results using $k = 20$ furnish $\Delta^{\mbox{\tiny (fluc)}} \sim\ln{(\sqrt{2})}$ for \textit{all maps} (large kicking intensities in the SMs). 
\end{abstract}
\cbeims{}
\maketitle

\textit{Introduction.} The interest in the quantum-classical correspondence of classically chaotic systems renewed in the last years due to the conjecture that puts a bound on the exponential growth rate  {$\Lambda\le 2\pi T$ ($T$ is the temperature)} of an out-of-time-ordered four-point correlator (OTOC) \cite{mal16}. Introduced in the context of the theory of superconductivity \cite{lar69}, $\Lambda$ is closely associated with the largest positive asymptotic Lyapunov exponent (LE) of the classical chaotic system for times shorter than the Ehrenfest time ($t<t_E$), for which quantum interference effects did not have time to become relevant.  Besides serving as a tool to understand the fundamentals in the quantum-classical relation of classically regular  \cite{buni19,lea20,wata20}, quasi-regular \cite{roz17, tomer20}
and chaotic \cite{roz17,mata18} systems, the behaviour of the OTOC attracted considerable attention in  many-body systems \cite{swi16,swi2,mal16-2,cotler18,fan17,knap17,altland17,urbina18,lea19} and in experiments \cite{rey17,fan18,way18,lea17}. In this context, we refer the readers to the recent interesting review about semiclassical many-body quantum chaos \cite{urbina22}.

In general, it is known that even though related, $\Lambda$ obtained from classical and quantum OTOCs and the LE ($\lambda$) are not precisely equal due to distinct order of averaging. {While $\Lambda$  is proportional to 
$\ln{(\mathbb{E}\left[ X\right])}$,  $\lambda$ is proportional to $\mathbb{E}\left[\ln{(X)}\right]$,
where $X=\{x_1, x_2,\ldots, x_N\}$ is a random variable related to the local finite-time  Lyapunov exponent (FTLE) 
$\lambda^{(t)}$, calculated at time $t$ (for $t\to\infty$, $\lambda^{(\infty)}=\lambda $).} Here 
$\mathbb{E}[\cdot]$ is the average over all points in the phase space.
In some cases, the relation between both is written as  $\Lambda=\lambda+\Delta$. For example, in the completely chaotic region of the kicked rotator it was observed numerically that $\Delta\approx \ln{\sqrt{2}}$ \cite{roz17} and 
in the many-body Dick model, $\Delta\approx 0.015$ \cite{lea19}.
{Even though the distinct order of averaging seems to be a mere mathematical 
property, it has deep interesting physical consequences. The mathematical background of our findings lies in  Jensen's inequality (JI)
$\mathbb{E}\left[\varphi{(X)}\right]\ge\varphi{(\mathbb{E}\left[ X\right])}$, where $\varphi(X)$ is a convex function.
Equality is restored when all variables in $X=\{x_1, x_2,\ldots,x_N\}$ are equal or when higher moments of $X$ are taken into account, which is the case considered here.}

This Letter demonstrates that  fluctuations of the 
FTLEs lead to the distinction between {$\Lambda$ and $\lambda$. 
Fluctuations of the local one-step FTLEs are well-known properties in dynamical systems  (see, for example, \cite{fuji83, celia04}}). Consequently, our results establish that for times $t<t_E$,  the classical and quantum exponential growth rates $\Lambda$ contain 
features of classical fluctuations not visible in the asymptotic $\lambda$ itself.  
{Analytical and numerical} results are shown for the chaotic logistic and tend maps and the classical and quantum standard maps.  The classical OTOC is defined as 
\begin{equation} 
\begin{split}
C_{\mbox{\tiny cl}}(t) = \mathbb{E}\left[ \left\{ x(t),p(0)\right\}^2  \right] =
	\mathbb{E}\left[ \left(\frac{\partial p(t) }{\partial x(0)}\right)^2  \right].
\end{split}
\label{CL1}
\end{equation}
\noindent
For chaotic systems it is expected that  $C_{\mbox{\tiny cl}}(t)\sim e^{2\Lambda t}$ and the exponential growth rate is determined through the numerical computation of $\Lambda=1/2\lim_{t\to\infty}\lim_{\Delta x(0)\to 0} \ln{[C_{\mbox{\tiny cl}}(t+1)}/C_{\mbox{\tiny cl}}(t)]$, where $\Delta x(0)$ are small initial displacements.  For one-dimensional systems $X=J(t)^2=e^{2t\lambda^{(t)}}$, where $J(t)$ is the Jacobian at time $t$. Applying JI  to $-\ln{(X)}$ (convex) we have $\ln{(\mathbb{E}\left[ X\right])}\ge \mathbb{E}\left[\ln{(X)}\right]$,  which provides the upper bound for the LE, namely  that $\lambda\le\Lambda$.

\textit{Chaotic logistic map} (LM). The map is defined as $x_{n+1} = 4\,x_n(1-x_n)$, with 
discrete times $n=1,2,\ldots$. The LM  has an invariant density  
$\rho_{\mbox{\tiny LM}}(x) = 1/(\pi\sqrt{x(1-x)})$ and the asymptotic LE is 
$\lambda_{{{\tiny\mbox{LM}}}} = \ln{(2)}$. Using the definition of the OTOC for 
the map, we have
\begin{eqnarray} 
C^{\mbox{\tiny LM}}_{\mbox{\tiny cl}}(n) =	\mathbb{E}\left[ \left(\frac{\partial x_{n} }{\partial x_0}\right)^2  \right]=
\mathbb{E}\left[ \left(e^{2\sum_{n}\lambda^{(n)}_{{{\tiny\mbox{LM}}}}}\right)  \right],\label{CLlog}\\
\nonumber
\end{eqnarray}
where $\lambda^{(n)}_{{{\tiny\mbox{LM}}}}=\ln{|J_n|}$ are the local \textit{one-step} FTLEs and $J_n = {\partial x_n }/{\partial x_{n-1}}=4-8x_{n-1}$ is the Jacobian of the map at time $n$. {Note that ${\partial x_{n} }/{\partial x_0}=J_nJ_{n-1}\ldots J_2J_1$.} {The local one-step FTLEs are fluctuating quantities and are responsible for the emergence of a nontrivial probability density, as observed previously for the LM \cite{celia04}}, that asymptotically converges 
to a delta centered at $\lambda_{{{\tiny\mbox{LM}}}} \left(=_{lim_{N\to\infty}}1/N\sum_n^N \lambda^{(n)}_{{{\tiny\mbox{LM}}}}\right)$. Thus, fluctuations of the FTLEs are expected to be relevant when determining Eq.~(\ref{CLlog}) \footnote{For the present analysis it is not adequate to use 
$\mathbb{E}\left[ \left(e^{2\sum_{n}\lambda^{(n)}_{{{\tiny\mbox{LM}}}}}\right)  \right]= \mathbb{E}\left[\left(e^{2N \lambda_{{{\tiny\mbox{LM}}}}}\right)\right]$ due to the relevance of each one-step FTLE.}.

Concerning the term in the middle of Eq.~(\ref{CLlog}), it is easy to show  that
$C^{\mbox{\tiny LM}}_{\mbox{\tiny cl}}(1) =\int_0^1dx_0 \rho_{\mbox{\tiny LM}}(x_0) \left({\partial x_{1} }/{\partial 
x_0}\right)^2= 8,$ $ C^{\mbox{\tiny LM}}_{\mbox{\tiny cl}}(2)= 64, C^{\mbox{\tiny LM}}
_{\mbox{\tiny cl}}(3)=512, C^{\mbox{\tiny LM}}_{\mbox{\tiny cl}}(4)=4096,\ldots$. 
From this we  
obtain the \textit{exact} OTOC exponent $\Lambda^{\mbox{\tiny (exact)}}_{\mbox{\tiny LM}}=
\ln{[C_{\mbox{\tiny cl}}^{\mbox{\tiny LM}}(2)/C_{\mbox{\tiny cl}}^{\mbox{\tiny LM}}(1)]}/2=
\ln{[C_{\mbox{\tiny cl}}^{\mbox{\tiny LM}}(3)/C_{\mbox{\tiny cl}}^{\mbox{\tiny LM}}(2)]}/2=
\ln{[C_{\mbox{\tiny cl}}^{\mbox{\tiny LM}}(4)/C_{\mbox{\tiny cl}}^{\mbox{\tiny LM}}(3)]}/
2=3\ln{(2)}/2$. These are \textit{two} iterations processes, namely from $n=0\to n=2, n=1\to 
n=3$ and 
$n=2\to n=4$, respectively. 
Observe that $\Lambda^{\mbox{\tiny (exact)}}_{\mbox{\tiny LM}} = \lambda_{\mbox{\tiny LM}} + \ln{(\sqrt{2})}$, so that
$\Delta^{\mbox{\tiny (exact)}}_{{\tiny\mbox{LM}}}
=\ln{(\sqrt{2})}$ 
gives the exact gap between both exponents.

To demonstrate the fluctuation properties of the gap,  we use a numerical method to determine $\Delta^{\mbox{\tiny (fluc)}}_{{\tiny\mbox{LM}}}$.
Motivated by the  exact results obtained above for the \textit{two iterations} processes in the determination of the middle term from Eq.~(\ref{CLlog}), it is reasonable to use only two iterations to attain knowledge about the fluctuations of the one-step FTLEs which are
relevant to $C^{\mbox{\tiny LM}}_{\mbox{\tiny cl}}(n)$.  This simplifies enormously our task
since we only need two iterations for the calculation of the last term on the right of Eq.~(\ref{CLlog}). Thus,  in terms of the one-step FTLEs, we have 
\begin{eqnarray} 
&\Lambda_{\mbox{\tiny LM}}^{\tiny (1\to 2)} =\frac{1}{2} \ln{\mathbb{E}
\left[\left(\frac{\partial x_{2} }{\partial x_{0}}\right)^2 \right]} 
-\frac{1}{2}\ln{\mathbb{E}
\left[\left(\frac{\partial x_{1} }{\partial x_0} \right)^2  \right]},\cr
&  \cr
&\cr
= &\frac{1}{2}\ln{\mathbb{E}
\left[\left(e^{2\left(\lambda^{(2)}_{{{\tiny\mbox{\tiny LM}}}}+\lambda^{(1)}_{{{\tiny\mbox{\tiny LM}}}}\right)  }\right) \right]} 
-\frac{1}{2}\ln{\mathbb{E}
\left[\left(e^{2\lambda^{(1)}_{{{\tiny\mbox{\tiny LM}}}}}\right)  \right]}. \label{CLlogN}\\
\nonumber
\end{eqnarray}
It crucial to realize that $\lambda^{(1)}_{{{\tiny\mbox{\tiny LM}}}}$ and
$\lambda^{(2)}_{{{\tiny\mbox{\tiny LM}}}}$ are one-step FTLEs from $n=0\to 1$ and from $n=1\to 2$, respectively. In order to connect Eq.~(\ref{CLlogN}) with the fluctuations of the one-step FTLEs,
we  apply the generating function of the cumulants
\begin{eqnarray}
&\ln{\left(\mathbb{E}\left[\exp{\left(2\, \lambda^{(f)}_T\right)}\right]\right)} = \sum_{k=1}^{\infty}\tilde
\kappa^{(f)}_k\frac{(2)^k}{k!},\label{cum}
\label{cum}
\end{eqnarray}
to both terms in Eq.~(\ref{CLlogN}),  being 
$\lambda^{(f=2)}_T=\lambda^{(2)}_{{{\tiny\mbox{\tiny LM}}}}+\lambda^{(1)}_{{{\tiny\mbox{\tiny LM}}}}$ and $\lambda^{(f=1)}_T=\lambda^{(1)}_{{{\tiny\mbox{\tiny LM}}}}$,  
so that  $\tilde\kappa^{(f=2)}_k$ are the $k$-order cumulants of the sum $\lambda^{(2)}_{{{\tiny\mbox{\tiny LM}}}}+\lambda^{(1)}_{{{\tiny\mbox{\tiny LM}}}}$, and  $\tilde\kappa^{(f=1)}_k$ are the $k$-order cumulants of  $\lambda^{(1)}_{{{\tiny\mbox{\tiny LM}}}}$.  Note that using the 
one-step FTLEs, the time $n$ does not appear explicitly in Eq.~(\ref{cum}) since it is incorporated in the sum
$\lambda^{(2)}_{{{\tiny\mbox{\tiny LM}}}}+\lambda^{(1)}_{{{\tiny\mbox{\tiny LM}}}}$. Therefore
\begin{eqnarray} 
& \Lambda_{\mbox{\tiny LM}}^{(1\to 2)} 
 = \tilde\kappa^{(2)}_1- \tilde\kappa^{(1)}_1+\frac{1}{2}\sum_{k=2}^{\infty}
\left(\tilde\kappa^{(2)}_k -\tilde\kappa^{(1)}_k\right)\frac{(2)^k}{k !}, \label{cumk} 
\end{eqnarray}
{where $\tilde\kappa^{(1)}_k$ are the cumulants related to $\lambda^{(1)}_{{{\tiny\mbox{\tiny LM}}}}$, and  $\tilde\kappa^{(2)}_k$ are the cumulants related to the \textit{sum} $\lambda^{(2)}_{{{\tiny\mbox{\tiny LM}}}}+\lambda^{(1)}_{{{\tiny\mbox{\tiny LM}}}}$, 
namely the joint cumulants of $\lambda^{(2)}_{{{\tiny\mbox{\tiny LM}}}}$ and $\lambda^{(1)}_{{{\tiny\mbox{\tiny LM}}}}$ \footnote{Here we use the joint cumulant of the variables $\lambda^{(2)}_{{{\tiny\mbox{\tiny LM}}}}$ and $\lambda^{(1)}_{{{\tiny\mbox{\tiny LM}}}}$, namely $\tilde\kappa_1^{(2)} = \mathbb{E}\left[\lambda^{(1)}_{{{\tiny\mbox{\tiny LM}}}} \lambda^{(2)}_{{{\tiny\mbox{\tiny LM}}}} \right]-\mathbb{E}\left[\lambda^{(1)}_{{{\tiny\mbox{\tiny LM}}}}\right] \mathbb{E}\left[\lambda^{(2)}_{{{\tiny\mbox{\tiny LM}}}}\right] = \mathbb{E}\left[\lambda^{(1)}_{{{\tiny\mbox{\tiny LM}}}} \lambda^{(2)}_{{{\tiny\mbox{\tiny LM}}}} \right]-\lambda_{\mbox{\tiny LM}}^2$, and so on for higher moments.}}, so that
\begin{eqnarray}
& \Lambda_{\mbox{\tiny LM}}^{(1\to 2)} \sim
\left\{ \lambda_{\mbox{\tiny LM}}+ \left(
\mathbb{E} \left[ \lambda^{(1)}_{\tiny\mbox{\tiny LM}}\lambda^{(2)}_{\tiny\mbox{\tiny LM}}    
\right] - \lambda_{\mbox{\tiny LM}}^2 -\kappa_2^{(1)} \right)+\right.\cr
& \left. \right.\cr
&\left. \left( \frac{3}{2} \mathbb{E} \left[ \lambda^{(1)}_{\tiny\mbox{\tiny LM}}\left(\lambda^{(2)}_{\tiny\mbox{\tiny LM}}\right)^2 \right]  +\frac{3}{2}\mathbb{E} \left[ \left(\lambda^{(1)}_{\tiny\mbox{\tiny LM}}\right)^2\lambda^{(2)}_{\tiny\mbox{\tiny LM}} \right] \right.\right.\cr
& \left.\left.\right. \right.\cr
& \left.\left.- 2 \lambda_{\mbox{\tiny LM}}^3-\kappa_3^{(1)} \right)+
\ldots\right\},
\nonumber
\end{eqnarray}
where we used $\kappa^{(2)}_1=2\,\kappa^{(1)}_1=2\,\lambda_{\mbox{\tiny LM}}$ and $\tilde\kappa^{(1,2)}_k=\kappa^{(1,2)}_k n^k$ (with $n=1$).
Taking into account the first $k=20$ cumulants we determine (see below) 
\begin{equation}
\Lambda_{\mbox{\tiny LM}}^{(1\to 2)} \sim \left[ \lambda_{\mbox{\tiny LM}} + 0.347500\right]\label{fin}. 
\end{equation}
Thus, the contribution of the fluctuations  leads to  $\Delta^{\mbox{\tiny (fluc)}}_{{\tiny\mbox{LM}}}(20)\sim 0.347500$, which is close to the analytical gap $\Delta^{\mbox{\tiny (exact)}}_{{\tiny\mbox{LM}}}=\log{(\sqrt{2})}\sim 0.346574$. To determine $\Delta^{\mbox{\tiny (fluc)}}_{{\tiny\mbox{LM}}}(k)$ we integrate numerically the central moments $\Upsilon_{\mbox{\tiny LM}}^{k}(N)$ (see \footnote{The cumulants in terms of the central moments can be obtained from expansions of the incomplete Bell polynomials. As an example, the first terms are
$\kappa_2 n =  \Upsilon^{(2)}, \kappa_3 n^2 =  \Upsilon^{(3)}, $
$\kappa_4 n^3 =  \Upsilon^{(4)} - 3 (\Upsilon^{(2)})^2,\ldots, $
$\kappa_5 n^4 =  \Upsilon^{(5)} - 10 \Upsilon^{(3)}\Upsilon^{(2)},\ldots$. To not confuse
the readers we did not use $n=1$ here.})
\begin{eqnarray}
\Upsilon_{\mbox{\tiny LM}}^{k}(1)&= & \int_0^1 dx_n\,\rho_{\mbox{\tiny LM}}(x_n)\left[
\lambda^{(1)}_{\mbox{\tiny LM}}-\lambda_{\mbox{\tiny LM}} \right]^k,\cr
& &\cr
\Upsilon_{\mbox{\tiny LM}}^{k}(2)&= & \int_0^1 dx_n\,\rho_{\mbox{\tiny LM}}(x_n)\left[
\lambda^{(2)}_{\mbox{\tiny LM}}+\lambda^{(1)}_{\mbox{\tiny LM}}-2\lambda_{\mbox{\tiny LM}} \right]^k.\cr
\nonumber
\label{kappai2}
\end{eqnarray}
\begin{figure}[h!]
\centering
\includegraphics[scale=0.8]{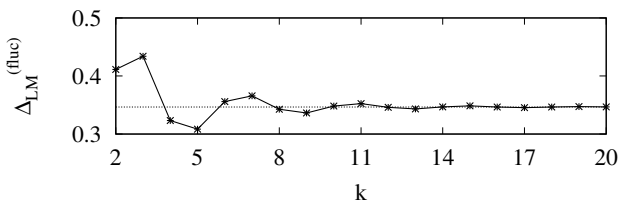}
\caption{Plotted is $\Delta^{\mbox{\tiny (fluc)}}_{\mbox{\tiny LM}}$ 
as a  function of the higher-order cumulants for the two iterations case.
Dashed line shows $\ln{(\sqrt{2})}$  for reference.}
 \label{fig1}
\end{figure}	
Results for  $\Delta^{\mbox{\tiny (fluc)}}_{{\tiny\mbox{LM}}}(k)$  are presented in Fig.~\ref{fig1} as a function of the cumulants' order $k$, and shows that it
converges to  $0.347500$. 
Thus, we expect that for $k\to\infty$ cumulants, the gap converges to $
\Delta^{\mbox{\tiny (fluc)}}_{{\tiny\mbox{LM}}}\to \log{\sqrt{2}}\sim
0.3465735$. Worth mentioning is that the cumulant expansion of the individual
terms in Eq.~(\ref{CLlogN}) increases without bounds, but the difference between 
them converges following Fig.~\ref{fig1}. Furthermore, Eqs.~(\ref{CLlogN}) and (\ref{cumk}) could be generalized to 
$\Lambda_{\mbox{\tiny LM}}^{(n\to n+1)}$, however,  the determination of $\Upsilon_{\mbox{\tiny LM}}^{k}(N)$ becomes harder and harder as $n$ increases and does not provide new relevant information.

\textit{Tend map} (TM). For the tend map, defined as $x_{n+1}= 2x_n$ for $x_n<1/2$,
and  $x_{n+1}= 2(1-x_n)$ for $x_n\geq 1/2$, the asymptotic LE is 
$\lambda_{\mbox{\tiny TM}}=\ln{(2)}$. The logistic and tend maps have the 
same LE \cite{Lichtenberg}. The OTOC growth rate is determined  exactly as
$\Lambda^{\mbox{\tiny (exact)}}_{\mbox{\tiny TM}}=\ln{[C_{\mbox{\tiny cl}}^{\mbox{\tiny TM}}(n)/C_{\mbox{\tiny cl}}^{\mbox{\tiny TM}}(1)]}/[2(n-1)]=\lambda_{\mbox{\tiny TM}}$. 
Since for the TM the invariant density is $\rho_{\mbox{\tiny TM}}(x)=1$,  the 
FTLEs are independent of the ICs, and no fluctuations are expected, so the 
corresponding central moments are exactly zero, leading  to 
$\Delta^{\mbox{\tiny (fluc)}}_{\mbox{\tiny TM}}=0$.  This trivial example establishes that when fluctuations of the FTLEs are absent, the OTOC and Lyapunov exponents are identical.

{\textit{Expressing the} OTOC \textit{in terms of} CLVs. Before discussing
results for the classical and quantum SMs, we present an expression for the classical OTOC em terms of CVLs in the two-dimensional continuous case.
We write  the right-hand side of Eq.(\ref{CL1})  as a function of quantities related to the evolution in the
 tangent space $T_\mathbf{x}M\equiv\mathbb{R}^2$, namely  in the CLVs basis, 
$\{v_\mathbf{x}\}=\{v_\mathbf{x}^{(u)}, v_\mathbf{x}^{(s)} \}$, which generate the Oseledec 
unstable $\{E_\mathbf{x}^{(u)}\}$ and stable  $\{E_\mathbf{x}^{(s)}\}$   subspaces with the properties
$D_\mathbf{x}f^t v_\mathbf{x}^{(i)} = \gamma_{i,\mathbf{x}}^{(t)} \ v_{i,\mathbf{x}+t}^{(t)},
\angle (E_{i,\mathbf{x}},E_{j,\mathbf{x}})\neq 0,  (\mbox{for}\, i\neq j)$ and $
\lim_{t\rightarrow\infty} \frac{1}{t}\log || D_\mathbf{x}f^t v_{i,\mathbf{x}}^{(t)} || = \lambda^{(\infty)}_{i}$ being the magnitude of the asymptotic Lyapunov exponent, 
with $i=u$ or $i=v$.  It is possible to show, after straightforward manipulation, that 
\begin{eqnarray}
C_{\mbox{\tiny cl}}^{\mbox{\tiny{(CLV)}}} (t) & =  
\mathbb{E}\left[f_{u,\mathbf{x}}^{(t)}\,f_{u,\mathbf{x}}^{(t)}\,e^{2t\lambda_{u,\mathbf{x}}^{(t)}} +
f_{s,\mathbf{x}}^{(t)}\,f_{s,\mathbf{x}}^{(t)}\,e^{2t\lambda_{s,\mathbf{x}}^{(t)}} \right]\
\cr
&\cr
 &\hspace*{-1cm}-  2 
\mathbb{E}\left[f_{u,\mathbf{x}}^{(t)}\,f_{s,\mathbf{x}}^{(t)}\,e^{t\left(\lambda_{u,\mathbf{x}}^{(t)}+\lambda_{s,\mathbf{x}}^{(t)}\right)}	\right],\label{CL3}\\
\nonumber
\end{eqnarray}
with the amplitudes\\
$f_{u,\mathbf{x}}^{(t)} = {\cos\left(\phi_\mathbf{x} + {\theta_\mathbf{x}}/
{2}\right)
	\cos\left(\phi_{f^t(\mathbf{x})} - {\theta_{f^t(\mathbf{x})}}/{2}\right)
	}/{\sin\theta_\mathbf{x}},$
$f_{s,\mathbf{x}}^{(t)}=
{\cos(\phi_\mathbf{x} - {\theta_\mathbf{x}}/{2})
	\cos\left(\phi_{f^t(\mathbf{x})} + {\theta_{f^t(\mathbf{x})}}/{2}\right)
	}/{\sin\theta_\mathbf{x}}.$
Equation (\ref{CL3}) furnishes explicitly the quantity $C_{\mbox{\tiny cl}}^{\mbox{\tiny{(CLV)}}}(t)$  as a function of: the finite-time CLV  $\lambda_{u,\mathbf{x}}^{(t)}$, related to the unstable manifold, the finite-time CLV  $\lambda_{s,\mathbf{x}}^{(t)}$  related to the stable manifold, the angle  $\theta_\mathbf{x}$ between both manifolds, their time derivative $\theta_{f^t(\mathbf{x})}$, the angle $\phi_\mathbf{x}$, which is the angle between $\theta_\mathbf{x}/2$ and the 
horizontal axis, and $\phi_{f^t(\mathbf{x})}$, its time derivative. We notice
that the CLVs $\lambda_{u,\mathbf{x}}^{(t)}$ and $\lambda_{s,\mathbf{x}}^{(t)}$ are 
calculated for finite times $t$, and only for $t\to\infty$ they lead to the usual asymptotic LEs 
$\lambda^{(\infty)}_{u}$ and $\lambda^{(\infty)}_{s}$, respectively. In other words, for short times, the local
values of $\lambda_{u,\mathbf{x}}^{(t)}$ and $\lambda_{s,\mathbf{x}}^{(t)}$, and their fluctuations, are essential for the behavior of the OTOC. Furthermore, the amplitudes of the exponents provide a clear contribution to the underline dynamics. For example, for ${\sin\theta_\mathbf{x}}\to0$  
an alignment between CLVs occurs and relevant contributions from the amplitudes of Eq.~(\ref{CL3}) are expected \footnote{More details of these contributions will be considered elsewhere.}. Recent works in other contexts focus on the role of prefactors to the OTOC \cite{kitaev19}.

\textit{The standard map} (SM).  The classical dissipative map is defined as \cite{chiri79} $p_{n+1} =  \gamma\,p_n + \frac{K}{2\pi}\sin{(2\pi q_n)} \mbox{(mod 1)},$ and $q_{n+1} =  q_n + p_{n+1} \mbox{(mod 1)}$,
where ($p_n,q_n$) are conjugate variables, $n=1,2,\dots$ the discrete time,
$\gamma$ is the dissipation parameter, and $K$ is the nonlinear parameter. For the analytical analysis of the SM, we use  Eq.~(\ref{CL1}) in the form
\begin{equation} 
C^{\tiny \mbox{\tiny SM}}_{\mbox{\tiny cl}}(n) =	\mathbb{E}\left[ \left(\frac{\partial p_{n} }{\partial q_0}\right)^2  \right] = \int_0^1\int_0^1 dq_0\, dp_0 \left(\frac{\partial p_{n} }{\partial q_0}\right)^2,
\label{CLlogSM}
\end{equation}
where the integration is over all phase-space initial conditions.\\ 
\textit{The conservative case} ($\gamma=1$). The analytical LE can be estimated from $\lambda^{\mbox{\tiny (exact)}}_{\tiny \mbox{SM}}=\int_0^1 dq\ln |L(q)|$, with $L(q)=1+ k(q)/2 + \mbox{sgn}[k(q)]\sqrt{k(q)(1+k(q)/4)}  $ and $k(q)=K\cos{(2\pi q)}$ \cite{chiri79}. Fluctuations of stability exponents in the SM have been already studied in another context \cite{tomso07}. It is known that for $K>4$ only one chaotic component lives in the phase space \cite{Lichtenberg}. Therefore,
for large values of  $K$, a completely chaotic motion is observed, and the asymptotic LE is $\lambda_{\tiny \mbox{SM}}=\ln{(K/2)}$.  Equation (\ref{CLlogSM}) furnishes  \textit{exact} expressions, namely 
$C^{\tiny \mbox{\tiny SM}}_{\mbox{\tiny cl}}(1) = {K^2/2}$ and 
$C^{\tiny \mbox{\tiny SM}}_{\mbox{\tiny cl}}(2) = K^2+{K^4/4}$,
so that
\begin{equation} 
 \Lambda^{(1\to 2)}_{\tiny \mbox{\tiny SM}}=\frac{1}{2}\ln{\left[\frac{C^{\tiny \mbox{SM}}_{\mbox{\tiny cl}}(2)}{C^{\tiny \mbox{\tiny SM}}_{\mbox{\tiny cl}}(1)}\right]}=-\ln{\sqrt{2}}+\frac{\ln{(4+K^2)}}{2}.
 \end{equation} 
Therefore, the gap is 
$\Delta^{\mbox{\tiny (exact)}}_{\tiny \mbox{SM}} =\Lambda^{(1\to 2)}_{\tiny \mbox{SM}}-\lambda_{\tiny \mbox{SM}}=\ln{\sqrt{2}}+\frac{1}{2}\ln{(4+K^2)}-\ln{(K)}$, which for  $K^2\gg 4$ reduces
to $\Delta^{\mbox{\tiny (exact)}}_{\tiny \mbox{SM}} \approx  \ln{(\sqrt{2})}$. Amazingly, this is the same gap $\Delta^{\mbox{\tiny (exact)}}_{\tiny \mbox{LM}} $ obtained for the LM,  which is a dissipative system.  We present  results for the SM with $K\ge 4$, since for smaller values of $K$ the dynamic is mixed (regular and chaotic), and the classical and quantum averages lead to additional difficulties which, besides being of general interest, are not essential for the goal of the present work.

For the numerical results, we initially compared the time evolution of 
$C_{\mbox{\tiny cl}}^{\mbox{\tiny{(CLV)}}} (n)$  from Eq.~(\ref{CL3}) with 
$C_{\mbox{\tiny cl}}(n)$ from Eq.~(\ref{CL1}), obtained using some 
small initial  displacements $\Delta x(0)$. Both results are in full agreement.
However, the $C_{\mbox{\tiny cl}}^{\mbox{\tiny{(CLV)}}} (n)$ from Eq.~(\ref{CL3}) is much superior in terms of the stability for longer iterations, since it does not depend on $\Delta x(0)$.  Thus,  to reckon the exponent $\Lambda^{\mbox{\tiny (CLV)}}_{\mbox{\tiny SM}}(n)$ we use  Eq.~(\ref{CL3}). To obtain the fluctuations of the CLVs, we determine numerically the distributions of $\lambda^{(t)}_{u,{\bold x}}$ at the Ehrenfest time\footnote{The Ehrenfest time depends on $K$. For details and specific values of the Ehrenfest times, we refer the reader to Ref.~\cite{roz17}.},  and obtain the $k=20$ first central moments directly from these distributions. Such higher moments lead again to $\Delta^{\mbox{\tiny (fluc)}}_{\tiny \mbox{SM}}$, whose convergences are similar to those obtained for the LM in Fig.~\ref{fig1}(a). Figure \ref{fig2}(a) shows results for  $\Delta^{\mbox{\tiny (fluc)}}_{\tiny \mbox{SM}}$ (blue squares),  $\Delta^{\mbox{\tiny (CLV)}}_{\tiny \mbox{SM}}= \Lambda^{\mbox{\tiny (CLV)}}_{\tiny \mbox{SM}}-\lambda^{\mbox{\tiny (exact)}}_{\tiny \mbox{SM}}$ (red crosses) for distinct values of $K$, together with the exact results  $\Delta^{\mbox{\tiny (exact)}}_{\tiny \mbox{SM}} $ (dark-pink squares).  For $K\ge 8$, $\Delta^{\mbox{\tiny (CLV)}}_{\tiny \mbox{SM}}$ and  $\Delta^{\mbox{\tiny (exact)}}_{\tiny \mbox{SM}} $ are indistinguishable,  and approach the value $0.346$ for $K=1000$. The values of  $\Delta^{\mbox{\tiny (fluc)}}_{\tiny \mbox{SM}}$ are very close, even though a bit smaller. The discrepancy between distinct curves for smaller values of $K$ is surely a consequence of the larger amount of dynamical fluctuations due to sticky motion\cite{Lichtenberg}, which strongly depend on appropriate averages.

Before we proceed to the quantum analysis, some information 
must be given. The  quantum OTOC is obtained 
numerically from  $C_{\mbox{\tiny Q}}(n) = \mathbb{E}\left\{\left[ \hat q(n), \hat p (0)\right]^2\right\}$, where ($\hat q,\hat p$) are the corresponding position and momentum operators, and $[,]$ denoting the commutator. 
The quantum SM problem is described using the kicked Hamiltonian operator (dimensionless units) $\hat{\cal{H}} =\hat{p}/2 +K/(4\pi^2)\cos{(2\pi\hat q)} \sum_{m=0}^{\infty}\delta (t-m\tau)$, and $C^{\mbox{\tiny (SM)}}_{\mbox{\tiny Q}}(n)$
is obtained from the numerical integration of the corresponding 
Schr\"odinger equation. The associated OTOC exponent is named 
$\Lambda^{\mbox{\tiny (Q)}}_{\tiny \mbox{SM}}$.
We use individual angular-momentum eigenstates $|\Psi(0)\rangle=\sum_{n=-\infty}^{\infty}a_{n}^{(0)}|n\rangle$ and Gaussian wave packets $a_{n}^{(0)}=\exp(-\frac{\hbar_{eff}^{2}(n-n_{0})^{2}}{2\sigma^{2}})$, where $n_{0}=\frac{p_{0}}{\hbar_{eff}}$. For the numerical integration we use $p_{0}=0$, $\sigma=4$ and $|\Psi\rangle$, represented in a finite basis of eigenstates $|n\rangle$, $n \in [-N, N-1]$. Functions of $\hat{p}$ are applied on this basis, and functions of $\hat{q}$ are applied in the Fourier-transformed representation. We use an adaptive grid with $2\hbar_{eff}N~\in ~[2^{12}, 2^{16}]$ 
\cite{roz17}.
\begin{figure}[h!]
\centering
\includegraphics[scale=0.6]{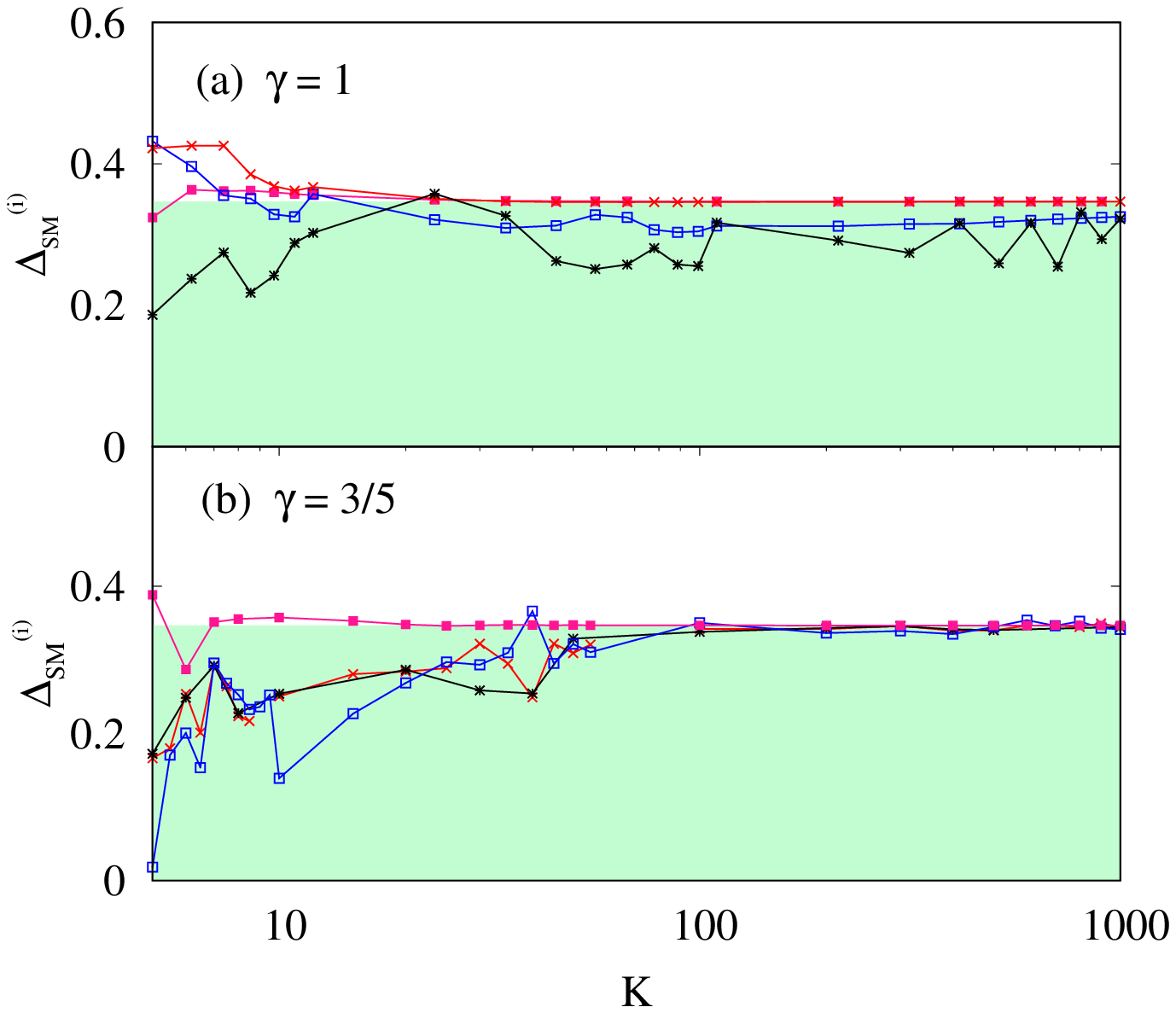}
\caption{Plotted is $\Delta_{\tiny \mbox{SM}}^{(i)} = \Lambda^{(i)}_{\tiny \mbox{SM}} - \lambda^{\mbox{\tiny (exact)}}_{\tiny \mbox{SM}}$ for the distinct calculated $\Lambda^{(i)}_{\tiny \mbox{SM}}$, namely the analytical result $\Lambda^{(i)}_{\tiny \mbox{SM}}= \Lambda^{(1\to 2)}_{\tiny \mbox{SM}}$ (black line), $\Lambda^{(i)}_{\tiny \mbox{SM}}= \Lambda^{\mbox{\tiny (CLV)}}_{\tiny \mbox{SM}}$ (red crosses), $\Lambda^{(i)}_{\tiny \mbox{SM}}= \Lambda^{\mbox{\tiny (fluc)}}_{\tiny \mbox{SM}}$  (blue square),  and $\Lambda^{(i)}_{\tiny \mbox{SM}}= \Lambda^{\mbox{\tiny (Q)}}_{\tiny \mbox{SM}}$ (black stars). (a) For the conservative case and (b) for the dissipative case. Upper borders of the light green rectangles show $\ln{(\sqrt{2})}$ for reference to be compared when $K$ is large.} 
\label{fig2}
\end{figure}
The quantum and classical OTOC exponents for the conservative SM 
were already considered recently \cite{roz17}, and our exponents
are in good agreement with those presented in Fig.~2 from \cite{roz17}, for the considered $K$ values. We choose not to repeat such a figure but, 
instead, display results for $\Delta_{\tiny \mbox{SM}}$, our main interest. Numerical results for $\Delta^{\mbox{\tiny (Q)}}_{\tiny \mbox{SM}}=\Lambda^{\mbox{\tiny (Q)}}_{\tiny \mbox{SM}}-\lambda^{\mbox{\tiny (exact)}}_{\tiny \mbox{SM}}$ as a function of $K$ are shown as black stars in Fig.~\ref{fig2}(a).  We notice that, except for specific values of $K$, $\Delta^{\mbox{\tiny (Q)}}_{\tiny \mbox{SM}}$ and $\Delta^{\mbox{\tiny (CLV)}}_{\tiny \mbox{SM}}$ are in relatively good agreement. Differences between both gaps are related to quantum averages and the number of eigenstates for each value of $K$. The determination of $\Lambda^{\mbox{\tiny (Q)}}_{\tiny \mbox{SM}}$ revealed to be a difficult numerical issue. Summarizing, Fig.~\ref{fig2}(a) demonstrates that the gaps between the distinct OTOCs and the classical LE are close to each other, and all quantities approach $\sim\ln{(\sqrt{2})}$ for large $K$ values, as accurately explained by the fluctuations of the finite-time CLVs.
 
\textit{The dissipative case} ($\gamma=3/5$). We could not obtain an analytical expression for the OTOC exponent using arbitrary values of $\gamma$. However, for $\gamma=3/5$, we attain
\begin{eqnarray}
\Lambda^{(1\to 2)}_{\tiny \mbox{SM}}=&\frac{1}{2}\log \Bigg{\{} \frac{34}{25} + \frac{K^{2}}{2} - \frac{\sqrt{5-\sqrt{5}}  \Big{(} K J_{0}(K) - J_{1}(K) \Big{)}}{ K \pi \sqrt{2}}  \cr & +\frac{5\sqrt{5-\sqrt{5}} \Big{(} K J_{0}(2K) + (K^{2}-1) J_{1} (2K)\Big{)}}{48 K \pi \sqrt{2}}\Bigg{\}},
\nonumber
\end{eqnarray}
with $J_i(K)$ ($i=0,1$) being the Bessel function of first type. Other specific values of $\gamma$ could be used. Furthermore,  Eq.~(\ref{CL3})  is used to reckon  $\Lambda^{\mbox{\tiny (CLV)}}_{\tiny \mbox{SM}}$, and the  FTLE $\lambda^{\mbox{\tiny (num)}}_{\tiny \mbox{SM}}$ is determined numerically, as usual.  Dissipation in the quantum model is introduced between the kicks by
coupling the main system to a zero-temperature environment. The density 
operator is determined as an ensemble mean over pure states obtained
from the quantum state diffusion \cite{gisin92} Ito-stochastic Schr\"odinger 
equation $|d\psi\rangle=-\hat{\cal H} |\psi\rangle dt +\sum_k 
(L_k-\langle L_k\rangle) |\psi\rangle d\xi_k-1/2\sum_k 
(L_k^{\dagger}L_k-2\langle L^{\dagger}_k\rangle)L_k+|\langle\psi\rangle |^2
 |\psi\rangle dt$. $L_k$ are the Lindblad operators with $k=1,2$ and 
 $\langle.\rangle$ stands for the expectation value. The Lindblad operators
 induce a damping $-\nu\langle\hat p\rangle$, and the dissipation parameter
 becomes $\gamma=e^{-\nu\tau}$, where the kicking time $\tau=\hbar_{\mbox{\tiny eff}}$ is the effective Planck's constant \cite{gisin92}. For details, we refer to \cite{martin16}. Figure \ref{fig2}(b) summarizes our results for the dissipative case with $\gamma=3/5$.  Plotted is $\Delta_{\tiny \mbox{SM}}^{\mbox{\tiny (i)}} = \Lambda^{\mbox{\tiny (i)}}_{\tiny \mbox{SM}} - \lambda^{\mbox{\tiny (num)}}_{\tiny \mbox{SM}}$ for the distinct calculated $\Lambda^{\mbox{\tiny (i)}}_{\tiny \mbox{SM}}$, namely the analytical result $\Lambda^{\mbox{\tiny (i)}}_{\tiny \mbox{SM}}= \Lambda^{\mbox{\tiny (exact)}}_{\tiny \mbox{SM}}$ (dark-pink  squares), $\Lambda^{\mbox{\tiny (i)}}_{\tiny \mbox{SM}}= \Lambda^{\mbox{\tiny (CLV)}}_{\tiny \mbox{SM}}$ (red crosses), $\Lambda^{\mbox{\tiny(i)}}_{\tiny \mbox{SM}}= \Lambda^{\mbox{\tiny (fluc)}}_{\tiny \mbox{SM}}$  (blue square),  and $\Lambda^{\mbox{\tiny (i)}}_{\tiny \mbox{SM}}= \Lambda^{\mbox{\tiny (Q)}}_{\tiny \mbox{SM}}$ (black stars). As for the conservative case, all quantities lead to a  gap $\Delta_{\tiny \mbox{SM}}\sim \ln{(\sqrt{2})}$ for larger $K$ values, nicely explained by the fluctuations of the one-step finite-time CLVs.

\textit{Conclusions.} Time fluctuations of the one-step FTLEs in the LM and the one-step finite-time CLVs in the SM are demonstrated to be the origin of the distinction between the classical and quantum OTOC exponential growth rate ($\Lambda$) and the classical LE ($\lambda$). The fluctuations are  quantified by higher-order cumulant expansions corrections $\Delta^{\mbox{\tiny (fluc)}}$, so that the upper bound $\Lambda=\lambda+\Delta^{\mbox{\tiny (fluc)}}$ is reached. Comparing the LM, and the SM for $K\ge 4$, the correction is $\Delta^{\mbox{\tiny (fluc)}}_{\mbox{\tiny LM}}\sim \Delta^{\mbox{\tiny (fluc)}}_{\mbox{\tiny SM}}\sim\ln{(\sqrt{2})}$.  Such approximated equality is intriguing: the statistical properties of the one-step FTLEs from the dissipative chaotic attractor of the LM are equal to those (CLVs) of the chaotic component of the conservative and dissipative SMs.  For the tend map, no fluctuations of the FTLEs are observed, leading to $\Delta^{\mbox{\tiny (fluc)}}_{\mbox{\tiny TM}}=0$. Thus, the quantum-classical correspondence regarding the exponential growth of instabilities in the SMs, becomes clear and uniquely described for  $t<t_E$ when taking into account the dynamical fluctuations of the one-step finite-time CLVs.  On account for the fact that the tend, logistic and standard maps are paradigmatic models describing a huge number of dynamical systems in distinct physical contexts, we are confident that the finite-time Lyapunov fluctuations producing the gap $\Delta^{\mbox{\tiny (fluc)}}$ should be a generic property.  Finally, it would be interesting to investigate the gap in many-body systems.

\acknowledgments{
MAPR and  MWB thank CNPq  (Brazil) for financial support (grant number from MWB is 310792/2018-5). The authors thank Prof.~C.~Anteneodo for suggestions about convergence of cumulants and Prof.~J.~D.~Urbina for discussions regarding the OTOC exponent in many-body systems.}


\end{document}